\documentstyle[11pt,epsfig]{article}
\title{\bf On extra forces from large extra dimensions}
\author{S. Jalalzadeh\thanks{email: s-jalalzadeh@sbu.ac.ir}, B. Vakili\thanks{email:
b-vakili@sbu.ac.ir} and H. R. Sepangi\thanks{email:
hr-sepangi@sbu.ac.ir}
\\ {\small Department of Physics,  Shahid Beheshti University, Evin, Tehran 19839, Iran}}
\begin{document}
\maketitle 

The motion of a classical test particle moving on a 4-dimensional
brane embedded in an $n$-dimensional bulk is studied in which the
brane is allowed to fluctuate along the extra dimensions. It is
shown that these fluctuations produce three different  forces
acting on the particle, all stemming from the effects of extra
dimensions. Interpretations are then offered to describe the
origin of these forces and  a relationship between the 4 and
$n$-dimensional mass of the particle  is obtained by introducing
charges associated with large extra dimensions.\vspace{5mm}

\section{Introduction}

Higher-dimensional gravity has a long history. This area of
research  began with the works of Kaluza and Klein  (KK)
\cite{1,2} where the inclusion of one compact extra dimension in
the space-time metric succeeded in unifying electromagnetism with
gravity. The metric in this scenario involves the fields $A_\mu$,
resulting in vacuum five dimensional Einstein equations which turn
out to include Maxwell equations with $A_\mu$ as the vector
potential describing electromagnetic fields. This unification can
also be extended to include Yang-Mills interactions. The extra
dimensions in KK approach have compact topology with
compactification scale $l$ so that at scales much larger than $l$
the extra dimensions are not observable. In more recent times,
theories in particle physics and gravity such as superstring,
which is a candidate for quantum gravity, supergravity and
M-theory have all used extra dimensions in their formalism with
different approaches to unification \cite{3,4}. In these theories
matter fields live in a 4-dimensional submanifold (brane) embedded
in a higher dimensional manifold (bulk) through which only
gravitons may propagate and extra dimensions are invisible at low
energy scales. As an example, in Horava-Witten's M-theory
\cite{horava} the standard model of interactions contained in the
$E_8 \times E_8$ heterotic string theory is confined to a 3-brane
but gravitons propagate in the 11-dimensional bulk.

The assumption that our universe might be embedded as a brane in a
1+3+$d$-dimensional bulk space has opened a new window to
cosmology. A number of papers have studied the cosmological
consequences of the embedded brane in a higher dimensional bulk.
The resulting modified Friedmann equations may be considered as
the main result in this context, see for example \cite{5}. The
bulk space in such theories can be considered as a
$4+d$-dimensional flat space \cite{6} in which a $4$-dimensional
brane is embedded. Also, non flat bulk spaces are studied by
Randall and Sundrum \cite{7,8}. A crucial difference between these
theories and standard KK models is that in the former  extra
dimensions are not necessarily compact and may be large.

As mentioned above, according to the brane world scenario
particles are confined to $4$-dimensional submanifolds of a
$4+d$-dimensional manifold. One of the important results of this
model are the extra forces that appear in the equations of the
particle motion. When the geodesic equation for a free particle in
$n$ dimensions is projected out onto a 4-dimensional (embedded)
world, these force-like terms appear on the right hand side of the
geodesic equations. These extra forces whose origin can be traced
to extra dimensions, may be interpreted in different manners. In
\cite{9} the nature of such forces in both compactified and
non-compactified extra dimensions is studied and shown not to
contradict physics in 4$D$. Comparison between different
approaches to the fifth force is undertaken in \cite{10,11}, where
interpretations for null and non-null bulk geodesics in the
context of brane theory, space-time-matter and other non-compact
approaches are developed. Also the relationship between $5D$ and
$4D$ masses of particles are obtained in \cite{12} with the help
of introducing an extra dimensional charge and shown to resemble
the mass formula $m=m_{0}/(1-v^2)^{1/2}$ in special relativity.

In all the above and other works \cite{youm} in this context,
extra forces are studied when a $4D$ brane is embedded in a $5D$
bulk. If the number of extra dimensions are greater than one  new
effects appear which are the subject of study in this paper. It is
important to note that the model we shall consider here is an
embedding model with large extra dimensions which  is not
equivalent to KK models \cite{13}. Indeed,  it is shown in
\cite{13} that KK gravity is not an embedding theory.

The structure of the paper is as follows: after a brief review of
geometry in section 2 we derive the equations of motion for a free
test particle in section 3. Projecting these equations onto  $4D$,
we obtain the $4D$ equations of motion in which extra forces
appear. In section 4 we consider an $n$-dimensional constant mass,
$M$, and obtain a relationship between this mass and a $4D$
observable mass, $m$. Introducing extra charges associated with
the test particle is done in section 5 together with addressing
the variations of masses and charges. Finally, in section 6 we
summarize and discuss the results.
\section{Geometrical setup}
In this section we briefly review the embedding procedure used in
the present work and originally appeared in \cite{14}. Consider
the background manifold $\overline{V}_{4}$ isometrically embedded
in ${V}_{n}$ by a map ${\cal E}: \overline{V}_{4}\rightarrow
V_{n}$ such that
\begin{equation} \label{A}
{\cal G}_{AB} {\cal Y}^{A}_{, \mu} {\cal Y}^{B}_{,
\nu}=\bar{g}_{\mu\nu} , \hspace{.5 cm} {\cal G}_{AB}{\cal
Y}^{A}_{, \mu}{\cal N}^{B}_{a}=0 , \hspace{.5 cm} {\cal
G}_{AB}{\cal N}^{A}_{a}{\cal N}^{B}_{b}=g_{ab}=\epsilon_{a};
\hspace{.5 cm} \epsilon_{a}=\pm 1
\end{equation}
where ${\cal G}_{AB}$ $(\bar{g}_{\mu\nu} )$ is the metric of the
bulk (brane) space $V_{n} (\overline{V}_{4})$ in arbitrary
coordinates, $\{{\cal Y}^{A}\}$ $(\{x^{\mu}\})$ are the basis of
the bulk (brane) and ${\cal N}^{A}_{a}$ are $(n-4)$ normal unit
vectors orthogonal to the brane. The local embedding is
constructed in a neighborhood of each point of the brane-world,
defining an embedding bundle whose total space consists of all
embedding spaces. The embedding equations are then derived from
the curvature tensor for each local embedding space, written in
the Gaussian frame defined by the embedded submanifold and the
normal vectors \cite{15}. From the point of view of brane-worlds,
this amounts to having a dynamic bulk whose geometry depends on
that of the brane-world as opposed to static or rigid embedding.
Since test particles are not exactly confined to a fixed brane, we
should analyze embedding of the neighboring perturbed branes. Note
that use of Nash's perturbative embedding procedure \cite{16}
guarantees that the perturbed branes are also isometrically
embedded.  Perturbation of $\overline{V}_{4}$ in a sufficiently
small neighborhood of the brane along an arbitrary transverse
direction $\xi$ is given by
\begin{equation} \label{B}
{\cal Z}^{A}(x^{\mu},\xi^{a})={\cal Y}^{A}+({\cal L}_{\xi}{\cal
Y})^A,
\end{equation}
where ${\cal L}$ represents the Lie derivative. By choosing
$\xi^{\mu}$ orthogonal to the brane we ensure gauge independency
\cite{14} and have perturbations of the embedding along a single
orthogonal extra direction $\bar{{\cal N}}_{a}$, giving the local
coordinates of the perturbed brane since
\begin{equation} \label{C}
{\cal Z}^{A}_{, \mu}(x^{\nu},\xi^{a})={\cal Y}^{A}_{,
\mu}+\xi^{a}\bar{{\cal N}}^{A}_{a,\mu}(x^{\nu}),
\end{equation}
where $\xi^{a}(a=5,...,n)$ are small parameters along ${\cal
N}^{A}_{a}$ parameterizing the extra noncompact dimensions. Also
one can see from equation (\ref{B}) that since the vectors
$\bar{{\cal N}}^{A}$ depend only on the local coordinates $x^\mu$,
they do not propagate along extra dimensions
\begin{equation} \label{D}
{\cal N}^{A}_{a}(x^{\mu})=\bar{{\cal
N}}^{A}_{a}+\xi^{b}[\bar{{\cal N}}_{b},\bar{{\cal
N}}_{a}]^{A}=\bar{{\cal N}}^{A}_{a}.
\end{equation}
The above assumptions lead to the embedding equations of the
perturbed geometry
\begin{equation} \label{E}
{\cal G}_{\mu\nu}={\cal G}_{AB}{\cal Z}^{A}_{, \mu}{\cal Z}^{B}_{,
\nu}, \hspace{.5 cm} {\cal G}_{\mu a}={\cal G}_{AB}{\cal Z}^{A}_{,
\mu}{\cal N}^{B}_{a}, \hspace{.5 cm} {\cal G}_{AB}{\cal
N}^{A}_{a}{\cal N}^{B}_{b}={\cal G}_{ab}.
\end{equation}
If we set ${\cal N}^{A}_{a}=\delta^{A}_{a}$, the metric of the
bulk space can be written in the following matrix form (Gaussian
frame)
\begin{equation} \label{F}
{\cal G}_{AB}=\left(\!\!\!\begin{array}{cc} g_{\mu\nu}+A_{\mu
c}A_{\nu}^{c} & A_{\mu a} \\ A_{\nu b} & g_{ab}
\end{array}\!\!\!\right),
\end{equation}
where
\begin{equation} \label{G}
g_{\mu\nu}=\bar{g}_{\mu\nu}-2\xi^{a}\bar{k}_{\mu\nu
a}+\xi^{a}\xi^{b}\bar{g}^{\alpha\beta}\bar{k}_{\mu\alpha
a}\bar{k}_{\nu\beta b},
\end{equation}
is the metric of the perturbed brane, so that
\begin{equation} \label{H}
\bar{k}_{\mu\nu a}=-{\cal G}_{AB}{\cal Y}^{A}_{,\mu}{\cal
N}^{B}_{a ; \nu},
\end{equation}
represents the extrinsic curvature of the original brane (second
fundamental form). Also, we use the notation $A_{\mu
c}=\xi^{d}A_{\mu cd}$, where
\begin{equation} \label{I}
A_{\mu cd}={\cal G}_{AB}{\cal N}^{A}_{d ; \mu}{\cal
N}^{B}_{c}=\bar{A}_{\mu cd},
\end{equation}
represents the twisting vector fields (normal fundamental form).
Any fixed $\xi^a$ signifies a new brane, enabling us to define an
extrinsic curvature similar to the original one by
\begin{equation} \label{J}
\widetilde{k}_{\mu\nu a}=-{\cal G}_{AB}{\cal Z}^{A}_{,\mu}{\cal
N}^{B}_{a ; \nu}=\bar{k}_{\mu\nu
a}-\xi^{b}\left(\bar{k}_{\mu\gamma a}\bar{k}_{\nu
b}^{\gamma}+A_{\mu ca}A^{c}_{b\nu}\right).
\end{equation}
Note that definitions (\ref{G}) and (\ref{J}) require
\begin{equation} \label{M}
\widetilde{k}_{\mu\nu a}=-\frac{1}{2}\frac{\partial {\cal
G}_{\mu\nu}}{\partial \xi^{a}}.
\end{equation}
In geometric language, the presence of gauge fields $A_{\mu a}$
tilts the embedded family of sub-manifolds with respect to the
normal vector ${\cal N} ^{A}$. According to our construction, the
original brane is orthogonal to the normal vector ${\cal N}^{A}$.
However,  equation (\ref{E})  shows that this is not true for the
deformed geometry. Let us change the embedding coordinates and set
\begin{equation}\label {re1}
{\cal X}_{,\mu }^{A}={\cal Z}_{,\mu }^{A}-g^{ab}{\cal
N}_{a}^{A}A_{b\mu }.
\end{equation}%
The coordinates ${\cal X}^{A}$ describe a new family of embedded
manifolds whose members are always orthogonal to ${\cal N}^{A}$.
In this coordinates the embedding equations of perturbed brane is
similar to the original one described by equations (\ref{re1}), so
that ${\cal Y}^{A}$ is replaced by ${\cal X}^{A}$. The extrinsic
curvature of a perturbed brane then becomes
\begin{equation}\label {re2}
k_{\mu \nu a}=-{\cal G}_{AB}{\cal X}_{,\mu }^{A}{\cal N}_{a;\nu }^{B}=\bar{k}%
_{\mu \nu a}-\xi ^{b}\bar{k}_{\mu \gamma a}\bar{k}_{,\,\,\nu b}^{\gamma }=-%
\frac{1}{2}\frac{\partial g_{\mu \nu }}{\partial \xi ^{a}},
\end{equation}%
which is the generalized York relation and shows how the extrinsic
curvature propagates as a result of the propagation of the metric
in the  direction of extra dimensions. In general the new
submanifold is an embedding in such a way that the geometry and
topology of the bulk space do not become fixed \cite{14}. We now
show that if the bulk space has certain Killing vector fields then
$A_{\mu ab}$ transform as the component of a gauge vector field
under the group of isometries of the bulk. Under a local
infinitesimal coordinate transformation for extra dimensions we
have
\begin{equation} \label{N}
\xi'^{a}=\xi^{a}+\eta^{a}.
\end{equation}
Assuming the coordinates of the brane are fixed, that is
$x'^\mu=x^{\mu}$ and defining
\begin{equation} \label{P}
\eta^{a}={\cal M}^{a}_{b}\xi^{b},
\end{equation}
then in the Gaussian coordinates of the bulk (\ref{F}) we have
\begin{equation} \label{O}
g'_{\mu a}=g_{\mu a}+g_{\mu b}\eta^{b}_{,a}+g_{ba}\eta^{b}_{,
\mu}+\eta^{b}g_{\mu a,b}+{\cal O}(\eta^2),
\end{equation}
hence the transformation of $A_{\mu ab}$ becomes
\begin{equation} \label{S}
A'_{\mu ab}=\frac{\partial g'_{\mu a}}{\partial
\xi'^{b}}=\frac{\partial g'_{\mu a}}{\partial \xi^{b}}-\eta^{A}_{,
b}\frac{\partial g'_{\mu a}}{\partial x^{A}}.
\end{equation}
Now, using $\eta^{a}_{, b}={\cal M}^{a}_{b}(x^{\mu})$ and
$\eta^{a}_{, \mu}={\cal M}^{a}_{b,\mu}\xi^{b}$ we obtain
\begin{equation} \label{R}
A'_{\mu ab}=A_{\mu ab}- 2A_{\mu c[a}{\cal M}^{c}_{b]}+{\cal
M}_{ab,\mu}.
\end{equation}
This is exactly the gauge transformation of a Yang-Mills gauge
potential. In our model the gauge potential can only be present if
the dimension of the bulk space is equal to or greater than six
$(n\geq 6)$, because the gauge fields $A_{\mu ab}$ are
antisymmetric under the exchange of extra coordinate indices $a$
and $b$. For example, let the bulk space be of Minkowskian type
with signature $(p,q)$. The tangent space of the brane has
signature $(1,3)$ implying that the orthogonal space has the
isometry group $ SO(p-1,q-3)$. Now, if ${\cal L}^{ab}$ is the Lie
algebra generators of the this group we have
\begin{equation} \label{Y}
[{\cal L}^{ab} , {\cal L}^{cd}]=C^{abcd}_{pq}{\cal L}^{pq},
\end{equation}
where $C^{abcd}_{pq}$ is the Lie algebra structure constants
defined by
\begin{equation} \label{W}
C^{abcd}_{pq}=2\delta^{[b}_{p}g^{a][c}\delta^{d]}_q.
\end{equation}
On the other hand if $F^{\mu\nu}=F^{\mu\nu}_{ab}{\cal L}^{ab}$ is
to be the curvature associated with the vector
potential$A^{\mu}=A^{\mu}_{ab}{\cal L}^{ab}$, we have
\begin{equation} \label{X}
F_{\mu\nu}=A_{\nu , \mu}-A_{\mu , \nu}+\frac{1}{2}[A_{\mu} ,
A_{\nu}],
\end{equation}
or in component form
\begin{equation} \label{V}
F_{\mu\nu ab}=A_{\nu ,\mu ab}-A_{\mu , \nu
ab}+\frac{1}{2}C^{mnpq}_{ab}A_{\mu mn}A_{\nu pq}.
\end{equation}
\section{Particle dynamics}
The equations of motion of a classical test particle moving on our
perturbed brane can be obtained from an action principle that
yields the $n$-dimensional geodesic equation as follows
\begin{equation} \label{a}
\frac{d^2{\cal Z}^{A}}{dS^2}+\Gamma^{A}_{BC}\frac{d{\cal
Z}^{B}}{dS}\frac{d{\cal Z}^{C}}{dS}=0,
\end{equation}
where $S$ is an affine parameter along the $n$-dimensional path of
the particle. To understand the nature of the motion in $4D$ we
must take $A=\mu$ in this equation and project it onto  our $4D$
brane. Also, with $A=a$ we obtain the motion in extra dimensions.
For this decomposition we need the Christoffel symbols of the bulk
space. Use of ${\cal G}_{AB}$ and its inverse
\begin{equation} \label{b}
{\cal G}^{AB}=\left(\!\!\!\begin{array}{cc} g^{\mu\nu} & -A^{\mu
a} \\ -A^{\nu b} & g^{ab}+A^{a}_{\alpha}A^{\alpha b}
\end{array}\!\!\!\right),
\end{equation}
results in
\begin{eqnarray}
\begin{array}{lllll}\vspace{.5 cm}
\bar{\Gamma}^{\mu}_{\alpha\beta} = \Gamma^{\mu}_{\alpha\beta} +
\frac{1}{2}\{ A_{\alpha c}F_{\beta}^{\,\,\,\mu c} + A_{\beta
c}F_{\alpha}^{\,\,\,\mu c} \} - \widetilde{k}_{\alpha\beta a}A^{\mu a}-\frac{1}{2}\{A_{c\alpha}A^c_{\,\,\,\,a\beta}+A_{c\beta}A^c_{\,\,\,\,a\alpha}\}A^{a\mu},\\
\vspace{.5 cm} \bar{\Gamma}^{\mu}_{\alpha a} =
-\widetilde{k}^{\mu}_{\,\,\,\,\alpha a} - \frac{1}{2}F^{\mu}_{\,\,\,\,\alpha a}-\frac{1}{2}\{A_{ab\alpha}A^{b\mu}+A^{\,\,\,\,\mu}_{ab}A^b_{\,\,\alpha}\},\\
\vspace{.5 cm} \bar{\Gamma}^{\mu}_{ab} = \bar{\Gamma}^{a}_{bc} =
0,\\ \vspace{.5 cm} \bar{\Gamma}^{a}_{\alpha\beta} =
\frac{1}{2}\left\{ \nabla_{\beta}A^{a}_{\,\,\,\,\alpha} +
\nabla_{\alpha}A^{a}_{\,\,\,\,\beta} +
A^{a\mu}A_{\,\,\,\,\alpha}^{c}F_{c\mu\beta} +
A^{a\mu}A_{\,\,\,\,\beta}^{c}F_{c\mu\alpha}\right\} +
k_{\,\,\,\,\alpha\beta}^{a} + A^{a}_{\,\,\,\,\mu}A^{\mu b}\widetilde{k}_{\alpha\beta b}-\\
\vspace{.5cm}\frac{1}{2}A^{a\mu}A_{b\mu}\{A_{c\alpha}A^{cb}_{\,\,\,\,\beta}
+A_{c\beta}A^{cb}_{\,\,\,\,\alpha}\},\\
\vspace{.5 cm} \bar{\Gamma}^{a}_{b \alpha} =
-\frac{1}{2}A^{a\beta}F_{b\alpha\beta} +
A^{a}_{\,\,\,\,b\alpha}+A^{a\mu}\widetilde{k}_{\alpha \mu
b}-\frac{1}{2}A^{a\mu}\{A_{bc\mu}A^c_{\,\,\alpha}+A_{bc\alpha}A^c_{\,\,\mu}\},
\end{array} \label{c}
\end{eqnarray}
where $\Gamma^{\mu}_{\alpha\beta}$ are the Christoffel symbols
induced on the perturbed brane. Substituting these relations into
(\ref{a}), the geodesic equation splits into the following
equations
\begin{equation} \label{d}
\frac{d^{2}x^{\mu}}{dS^2}+\Gamma^{\mu}_{\,\,\,\,\alpha\beta}\frac{dx^\alpha}{dS}
\frac{dx^\beta}{dS}={\cal Q}^{a}F^{\mu}_{\,\,\,\, \alpha
a}\frac{dx^\alpha}{dS}+2k^{\mu}_{\,\,\,\,\alpha
a}\frac{dx^\alpha}{dS}\frac{d \xi^a}{dS}+k_{\alpha\beta a}A^{\mu
a}\frac{dx^\alpha}{dS}\frac{dx^\beta}{dS},
\end{equation}
and
\begin{equation} \label{e}
\frac{d^{2}\xi^a}{dS^2}+\left(\nabla_{\alpha}A^{a}_{\beta}+A^{a
\mu}A^{c}_{\alpha}F_{c
\mu\beta}+k^{a}_{\alpha\beta}+A^{a}_{\mu}A^{\mu b}k_{\alpha\beta
b}\right)\frac{dx^\alpha}{dS}\frac{dx^\beta}{dS}+\left(2A^{a}_{\,\,\,\,\alpha
 b}-A^{a \beta}F_{b
\alpha\beta}\right)\frac{dx^\alpha}{dS}\frac{d \xi^b}{dS}=0,
\end{equation}
where
\begin{equation} \label{f}
{\cal Q}^a=\frac{d \xi^a}{dS}+A^{a}_{\beta}\frac{dx^\beta}{dS}=
\left(\frac{d\xi^a}{ds}+A^a_\beta\frac{dx^\beta}{ds}\right)\frac{ds}{dS}=Q^a\frac{ds}{dS},
\end{equation}
is a {\it charge}-like quantity (per unit mass) associated with
the particle and $ds$ is the line element of the $4D$ brane. It is
clear that this quantity appears because of the motion in extra
dimensions. From the point of view of a four dimensional observer
the above equations should rewritten in terms of an affine
parameter on the brane which we  take as $s$. With the help of
equations such as
$\frac{dx^{\mu}}{dS}=\frac{dx^{\mu}}{ds}\frac{ds}{dS}$ and
$\frac{d^2x}{dS^2}=\frac{d^2
x}{ds^2}(\frac{ds}{dS})^2+\frac{dx}{ds}\frac{ds}{dS}\frac{d^2s}{dsdS}$
and also use of (\ref{re2}) we can write
\begin{equation}\label{re3}
\frac{d^{2}x^{\mu}}{ds^2}+\Gamma^{\mu}_{\,\,\,\,\alpha\beta}\frac{dx^\alpha}{ds}
\frac{dx^\beta}{ds}= Q^{a}F^{\mu}_{\,\,\,\, \alpha
a}\frac{dx^\alpha}{ds}+2k^{\mu}_{\,\,\,\,\alpha
a}\frac{dx^\alpha}{ds}\frac{d \xi^a}{ds}+k_{\alpha\beta a}A^{\mu
a}\frac{dx^\alpha}{ds}\frac{dx^\beta}{ds}-k_{\alpha\beta
a}Q^a\frac{dx^{\alpha}}{ds}\frac{dx^{\beta}}{ds}\frac{dx^{\mu}}{ds},
\end{equation}
also equation (\ref{e}) simplifies to
\begin{equation} \label{g}
\frac{d
Q^a}{ds}=-\frac{dx^\alpha}{ds}\left(\widetilde{k}^{a}_{\,\,\,\,\alpha\beta}
\frac{dx^\beta}{ds}+A^{ a}_{\,\,\,\,\alpha b}Q^b
\right)-k_{\alpha\beta b}Q^bQ^a
\frac{dx^{\alpha}}{ds}\frac{dx^{\beta}}{ds}.
\end{equation}
Equation (\ref{re3}) is the $4D$ equation of motion and describes
the motion of a particle under the action of the force (per unit
mass)
\begin{equation} \label{h}
f^\mu=Q^aF^{\mu}_{\,\,\,\,\alpha a}\frac{dx^{\alpha}}{ds}+(2g^{\mu
\nu}-\frac{dx^{\mu}}{ds}\frac{dx^{\nu}}{ds})Q^ak_{a \alpha
\nu}\frac{dx^{\alpha}}{ds}+(k_{\alpha\beta a}A^{a
\mu}-2k^{\mu}_{\alpha
a}A^a_{\,\,\beta})\frac{dx^{\alpha}}{ds}\frac{dx^{\beta}}{ds}.
\end{equation}
The existence of this force arises from the motion of the test
particle not only in $4D$ but also along the extra dimensions. The
first term in (\ref{h}) is a {\it Lorentz}-like force (per unit
mass) the existence of which is expected since we showed in
section 2 that the twisting vectors $A_{\mu ab}$ transform as
gauge potentials in Yang-Mills theory. It should be noted that in
the case of only one extra dimension, this term disappears because
of the vanishing $F^{\mu\nu}$. However, having two extra
dimensions results in the usual electromagnetic force. A natural
extension would be to go to higher dimensions where one recovers
the Yang-Mills fields. Therefore, we may consider $Q^a$ as the
ratio of charge to mass of the test particle $Q^a=q^a/m$ where
$q^a$ and $m$ are the observable charge and mass of the particle
respectively. The second term in (\ref{h}) shows the forces coming
from the extrinsic curvature in the usual sense of the embedding
theories since we know that the extrinsic curvature measures the
amount that the embedded manifold is curved in the way it sits in
the bulk manifold. Finally, the third term is a force in which the
extrinsic curvature is coupled to gauge fields. We shall return to
this issue in section 5 and give a more comprehensive
interpretation of the second and third terms of the above
equation.
\section{Relationship between $4D$ and $nD$ masses}
In this section we obtain a relationship between the
$n$-dimensional constant mass $M$ and the effective $4D$
observable mass $m$ that an observer measures. Let us define the
momentum of the particle in terms of its $n$-dimensional velocity
as
\begin{equation} \label{i}
P^A=M U^A,
\end{equation}
where upon using $U^AU_A=1$ we have
\begin{equation} \label{j}
{\cal G}_{AB}P^{A}P^{B}=M^2.
\end{equation}
Note that Capital Latin indices are lowered and raised with the
$nD$ metric. To a $4D$ observer the motion is described by the
$4$-momenta $p_\mu$ such that
\begin{equation} \label{l}
g_{\mu\nu}p^{\mu}p^{\nu}=m^2.
\end{equation}
Here the Greek indices are lowered and raised by the brane metric.

In order to relate $m$ and $M$ we project the $n$-momentum $P^A$
onto  four dimensions. Assuming that this projection is done by
the vielbeins $e^{(M)}_{A}$ defined as ${\cal G}_{AB}=e^{(M)}_A
e^{(N)}_B \eta_{(M)(N)}$ where $\eta_{(M)(N)}$ is a block-diagonal
metric with $\eta_{(\mu)(\nu)}=g_{\mu \nu}$ and
$\eta_{(a)(b)}=g_{ab}$, it is easy to show that

\begin{equation} \label{m}
e^{(\alpha)}_{\mu}=\delta^{\alpha}_{\mu}, \hspace{.5 cm}
e^{(a)}_{\mu}=A^{a}_{\mu}, \hspace{.5 cm} e^{(\alpha)}_{a}=0,
\hspace{.5 cm} e^{(a)}_{b}=\delta^{a}_{b}.
\end{equation}
With the use of the above formulae and projections $P_A=e^{(M)}_A
p_M$ or $p^A=e^{(A)}_M P^M$, one obtains
\begin{equation} \label{n}
P_a=p_a, \hspace{.5 cm} P^a=p^{a}-A^{a}_{\alpha}p^{\alpha},
\hspace{.5 cm} P^\mu=p^\mu, \hspace{.5 cm} P_\mu=p_{\mu}+A_{\mu
a}p^a.
\end{equation}
Now equation (\ref{j}) can be rewritten as
\begin{equation} \label{o}
g_{\mu\nu}p^{\mu}p^{\nu}+g_{ab}p^{a}p^{b}=M^2,
\end{equation}
where we have made use of equations (\ref{b}) and (\ref{n}).
However, from (\ref{n}) we have
$P^a=p^{a}-A^{a}_{\mu}p^{\mu}=M\frac{d \xi^a}{dS}$, so we obtain
\begin{equation} \label{p}
p^a=M\frac{d
\xi^a}{dS}+mA^{a}_{\mu}\frac{dx^\mu}{ds}=m\left(\frac{d
\xi^a}{ds}+A^{a}_{\mu}\frac{dx^\mu}{ds}\right)=mQ^a,
\end{equation}
where we  have assumed $\frac{M}{dS}=\frac{m}{ds}$ \cite{11}.
Substituting this result in (\ref{o}) yields $m^{2}+m^{2}Q^{2}=M^
2$ or
\begin{equation} \label{s}
m=\frac{M}{\sqrt{1+Q^2}},
\end{equation}
where $Q^2=g_{ab}Q^a Q^b$. We may obtain this formula in an
alternative way. The bulk space metric can be written as
\begin{equation} \label{r}
dS^2=ds^2+g_{ab}\left(d \xi^{a}+A^{a}_{\mu}dx^{\mu}\right)\left(d
\xi^{b}+A^{b}_{\nu}dx^{\nu}\right).
\end{equation}
Dividing both sides of this equation by $ds^2$ and using
(\ref{f}), we find
$$\left(\frac{dS}{ds}\right)^2=1+g_{ab}Q^{a}Q^{b}.$$ Assuming
again that $\frac{dS}{M}=\frac{ds}{m}$, we are again led to
equation (\ref{s}).

Equation (\ref{s}) is an interesting relationship that can be
compared with  equation $m=\frac{m_0}{\sqrt{1-v^2}}$ in the
special theory of relativity, bearing in mind that here instead of
$v$ we have the velocity in extra dimensions because of the
appearance of $Q^2$. It should be noted that for a massless
particle in the bulk space ($M=0$ and $dS^2=0$), if all extra
dimensions are timelike, then from (\ref{s}) we have $m=0$, a
massless particle in the $4D$ brane. However, in the case of
having spacelike extra dimensions, the $4D$ mass depends on the
direction of the motion. If the particle does not undergo any
motion normal to the perturbed brane, then $d \xi^a+A^a_\mu dx^\mu
=0 $ and equation (\ref{r}) shows $ds^2=0$, {\it i.e.} we have a
massless particle on the brane. However, if the normal components
of the direction of motion are to be nonzero the particle moving
on the brane will be massive. These results are in agreement with
that of \cite{12} where the value of $m$ is calculated in a $5D$
cosmological model.

\section{Extra charges associated with particles}
Use of the above formalism in the case of $M\neq 0$ would now
enable us to introduce extra charges associated with the particle.
As is well known, when a charged particle is moving in an
electromagnetic field its Hamiltonian can be obtained from the
substitution  ${\bf p}\rightarrow {\bf p}+ e{\bf A}$, where $e$ is
the charge of the particle and ${\bf A}$ is the vector potential.
From this analogy and using the last equation of (\ref{n}) we
define the extra charges associated with the particle as follows
\begin {equation} \label{AB}
q^a=p^a=m Q^a.
\end{equation}
Now from this equation and equation (\ref{s}), it is clear that
\begin{equation} \label{AC}
dq^a=Q^adm+m dQ^a,
\end{equation}
and
\begin{equation} \label{AD}
dm=-\frac{1}{2} \frac{m}{1+Q^2}d(Q^2).
\end{equation}
Defining
\begin{equation} \label{AE}
d\omega^a=dQ^a-\frac{1}{2}Q^a\frac{d(Q^2)}{1+Q^2},
\end{equation}
equations (\ref{AC}) and (\ref{AD}) can be rewritten as
\begin{equation} \label{AF}
dq^a=m d\omega^a,
\end{equation}
\begin{equation} \label{AG}
dm=-g_{ab}q^b d\omega^a.
\end{equation}
The general solutions of these equations can be written in the
following form
\begin{equation} \label{AH}
m=M\cos(\lambda_a \omega^a),
\end{equation}
\begin{equation} \label{AI}
q^a=M \lambda^a\sin(\lambda_b \omega^b),
\end{equation}
where $\lambda^a$ are integration constants satisfying
$g_{ab}\lambda^a \lambda^b=1$ as a consequence of equation  $m^2 +
g_{ab}q^a q^b=M^2$.

Equations (\ref{AH}) and (\ref{AI}) show that the effective $4D$
mass and extra charges oscillate about the $n$-dimensional mass
$M$. Also these results show that we have not a conserved mass or
charge in a $4D$ brane, instead, the total mass $M$ is a conserved
quantity. Now, to realize the existence of the force terms on the
right hand side of equation (\ref{h}), we contract $f_{\mu}$ with
the four velocity $u^{\mu}=\frac{dx^{\mu}}{ds}$, with result
\begin{equation}\label{AJ}
f_{\mu}u^{\mu}=Q^ak_{a\alpha \beta}u^{\alpha}u^{\beta}-k_{a\alpha
\beta}A^a_{\mu}u^{\alpha}u^{\beta}u^{\mu}.\end{equation} On the
other hand from equations (\ref{AC}), (\ref{AD}) and definition of
$Q^a$ in (\ref{f}), one can show that
\begin{equation}\label{AM}
\frac{\dot{m}}{m}=Q^ak_{a\alpha
\beta}u^{\alpha}u^{\beta},\hspace{.5cm}k_{a\alpha
\beta}A^a_{\mu}u^{\alpha}u^{\beta}u^{\mu}=-\frac{Dq^a}{ds}A_{a\mu}u^{\mu},\end{equation}
where
\begin{equation}\label{AN}
\dot{m}=\frac{dm}{ds},\hspace{.5cm}\frac{Dq^a}{ds}=\frac{dq^a}{ds}+A^a_{\,\,\,b\alpha}q^bu^{\alpha}.\end{equation}
Substituting these results in equation (\ref{AJ}), we are led to
\begin{equation}\label{AO}
f_{\mu}u^{\mu}=\frac{\dot{m}}{m}+\frac{Dq^a}{ds}A_{a\mu}u^{\mu}.\end{equation}
We see that contrary to usual relativity for which
$f_{\mu}u^{\mu}=0$, here we have a non-zero value for this
quantity. Equation (\ref{AO}) shows that manifestations of extra
forces, as viewed by a $4D$ observer, are embodied in the
variation of inertial mass and charge of the particle. Note that
because of the antisymmetric property of $F^{\mu}_{\,\,\,\alpha
a}$ with respect to $\mu$ and $\alpha$, contraction of $u^{\mu}$
with the first term on the right hand side of equation (\ref{h})
is zero. Thus the two other terms are responsible for a non-zero
value of $f_{\mu}u^{\mu}$ and therefore of the mass and charge
variations. A similar result of variation of mass and spin due to
the extra forces in a $5D$ KK gravity is obtained in \cite{17}.
\section{Conclusions}
We have studied a model in which a $4D$ brane is embedded in an
$n$-dimensional bulk. In doing so, we have constructed a
non-compact KK model which is compatible with the embedding
hypothesis. This is in contrast to the works of other authors
\cite{9,10,11} where a non-compact KK metric in $5D$ is used and
the resulting model may not be considered as an embedding theory.

In so far as a test particle moving on such a bulk space is
concerned, a $4D$ observer would detect a $4D$ mass and Yang-Mills
charge stemming from the oscillations of the particle along
directions perpendicular to the brane. A force is also experienced
by the particle which again originates from the effects of extra
dimensions in the form of the extrinsic curvature and Yang-Mills
fields. We have seen that the existence of extra forces results in
the variations of mass and charge of the test particle as viewed by
the $4D$ observer. That is, the simplest interpretation of equation
(\ref{AJ}) in terms of $4D$ physics is to assume that the invariant
rest mass and charge of the particle may vary with respect to its
proper time $s$, due to the existence of higher dimensions.
Conversely, the observation of such a basic variation would indicate
the presence of an extra acceleration and this could come about
precisely because of the intrusion of the extra dimensions into the
$4D$ physics as indicated by equations (\ref{AJ}). If the bulk space
is $5D$, the particle only experiences the effects of the extrinsic
curvature. Such effects may be of interest in studying the classical
tests of general relativity in our solar system or in accounting for
the existence of dark matter in our galaxy in terms of geometrical
properties of the space-time.

\end{document}